\documentclass[acus]{jac2000} 
\usepackage{epsfig}
\def\gappeq{\mathrel{ \rlap{\raise.5ex\hbox{$>$}}
                      {\lower.5ex\hbox{$\sim$}}  } }
\def\lappeq{\mathrel{ \rlap{\raise.5ex\hbox{$<$}}
                      {\lower.5ex\hbox{$\sim$}}  } }
\begin{document}
\title{Ground Motion Model of the SLAC Site
\thanks{Work supported by the U.S. Department of Energy, 
Contact Number DE-AC03-76SF00515.}}
\author{Andrei~Seryi, Tor~Raubenheimer \\
{\it Stanford Linear Accelerator Center, Stanford University, 
Stanford, California 94309 USA}
}
\maketitle
\begin{abstract}
We present a ground motion model for the SLAC site. 
This model is based on recent ground motion studies performed at
SLAC as well as on historical data. The model includes wave-like, 
diffusive and systematic types of
motion. An attempt is made to relate measurable secondary 
properties of the ground motion with more basic characteristics
such as the layered geological structure of the surrounding earth, depth 
of the tunnel, etc.  This model is an essential step in evaluating
sites for a future linear collider.
\end{abstract}

\vspace{-.35cm}
\section{Introduction}
\vspace{-.15cm}
In order to accurately characterize the influence of ground motion 
on a linear collider, an adequate
mathematical model of ground motion has to be created. 
An adequate model would require an understanding of
the temporal and spatial properties of the motion and 
identification of the driving mechanisms of the motion. 
Eventually these must be linked to more general properties of a site 
like geology and urbanization density. 
In this paper, we consider one particular model 
based on measurements performed at the 
SLAC site \cite{fischer,ZDR,fftbwire,m95,slow1}. 
We use this model to illustrate existing methods of
modeling, as well as potential problems
and oversimplifications in the modeling techniques. 
In our particular case, the representation of the 
cultural noise, especially that generated 
inside the tunnel, is difficult to incorporate. However, the 
model provides a foundation 
to which many additional features can be added.

In general, the ground motion can be divided into `fast' and
`slow' motion. Fast motion ($f \gappeq $ a few Hz)
cannot be adequately corrected by a pulse-to-pulse feedback operating at
the repetition rate of the collider and 
therefore results primarily in beam offsets at the IP.
On the other hand, the beam offset due to slow motion can be compensated
by feedback and thus
slow motion ($f \lappeq 0.1$) results only in beam emittance growth.
Another reason to divide ground motion into fast and slow regimes is the
mechanism by which relative displacements are produced that 
appears to be different with
a boundary occuring around 0.1~Hz.  In the following, we will first describe the `fast'
motion and then we will present the `slow' motion which includes both
diffusive and systematic components.

\vspace{-.35cm}
\section{Fast Ground Motion}
\vspace{-.15cm}
Modeling of the ground motion requires knowledge of the 2-D
power spectrum $P(\omega,k)$. The fast motion is usually 
represented by quantities that can be measured directly: 
the spectra of absolute motion $p(\omega)$  
and the correlation $c(\omega, L)$ which shows 
the normalized difference in motion of two points 
separated by distance $L$.
The spectrum of relative motion $p(\omega,L)$ 
can be written 
as $p(\omega,L) = p(\omega) 2 (1-c(\omega, L))$
which in turn can be transformed into $P(\omega,k)$ \cite{sn}.

Measurements \cite{ZDR,vjlep} show that 
the fast motion in a reasonably quiet site 
consists primarily of elastic waves propagating 
with a high velocity $v$ (of the order of km/s).
The correlation is then completely defined by this 
velocity (which may be a function of frequency) 
and by the distribution of the noise sources. 
In the case where the waves propagate on the surface 
and are distributed uniformly in azimuthal angle, 
the correlation is given by 
$c(\omega,L)= \langle\cos(\omega L/v \, \cos(\theta))\rangle_{\theta}=
J_0 (\omega L/v)$ and the corresponding 
2-D spectrum of the ground motion is
$P(\omega,k)= 2 p(\omega) /\sqrt{(\omega/ v(f))^2-k^2} $, 
$|k| \le \omega/ v(f)$.

\begin{figure}[h]
\vspace{-.2cm}
\centering
{\vbox{
\epsfig{file=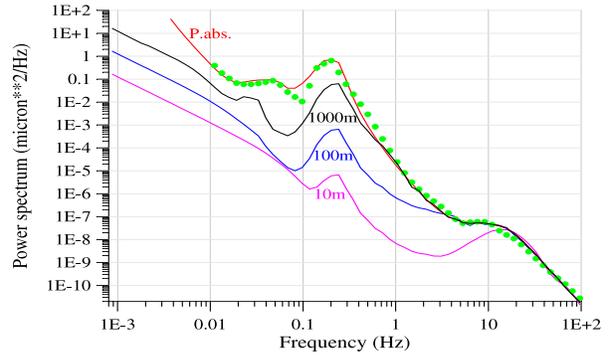,height=0.95\columnwidth,width=0.56\columnwidth,angle=-90}
}}
\vspace{-.3cm}
\caption{Measured \cite{ZDR} (symbols) and modeling 
spectra $p(\omega)$ 
of absolute motion and $p(\omega,L)/2$ spectra of relative motion 
for the 2~AM SLAC site ground motion model.}
\vspace{-.3cm}
\label{slacgm}
\end{figure}

The absolute power spectrum of the fast motion, 
assumed for the SLAC model, corresponds to measurements 
performed at 2~AM in one of the quietest locations at SLAC,
sector 10 of the linac \cite{ZDR},
(see Fig.\ref{slacgm}).
The spatial properties are defined by 
the phase velocity found from correlation measurements
$v(f) = 450+1900\exp(-f/2)$
(with $v$ in m/s, $f$ in Hz) \cite{ZDR}.

We believe that the frequency dependence of the measured phase velocity $v(f)$
is explained by the geological structure of the SLAC site 
where, as is typical, the ground rigidity and the density increase with depth.
The surface motion primarily consists of transverse waves whose phase velocity 
is given by $v_s\approx\sqrt{E/(2\rho)}$ and which are 
localized within one wavelength of the surface.
If one plots the quantity $v^2/\lambda$ 
versus wavelength $\lambda$, we 
see that this value is almost constant, 
varying from 
$3000$m/s$^2$
%
%
at $\lambda=100$m to $2000$m/s$^2$
%
%
at $\lambda=1000$m. This is consistent with a
ground density at the SLAC site that
ranges from $1.6\cdot 10^3$ within the upper 100~m
to $2.5\cdot 10^3$kg/m$^3$ at a kilometer depth and  
a Young's modulus $E$ which increases from $10^9$Pa at 100~m to
$10^{10}\,$Pa at 1000~m.
These results seem to be quite 
reasonable for the SLAC geology, and, as we will see 
below, they also agree with explanations of
the observed slow motion.
\\[2mm]

\vspace{-9mm}
\section{Slow Ground Motion}
Based on the arguments above, the wavelength at frequencies below 0.1~Hz 
quickly becomes much larger than the accelerator and eventually
exceed the earth's size.  In this case, the motion has little effect
on the accelerator and at some point the notion of waves is not really applicable.
Causes other than the wave mechanism must be responsible
for producing relative misalignments that are important
at low frequencies.  Such sources include the variation
of temperature in the tunnel, underground water flow,
spatial variation of ground properties combined with some
external driving force, etc. These causes can produce 
misalignments with rather short wavelength 
in spite of their low frequencies. 

The ATL model of diffusive ground motion \cite{atl}
is an attempt to describe all these complex
effects with a simple rule which states that the variance of 
the relative misalignment $\Delta X^2$ is proportional to 
a coefficient $A$, the time $T$ and the separation $L$: 
$\Delta X^2 = A T L$. In the spectral representation 
this rule can be written as 
$P(\omega,k) =A/ (\omega^2 k^2)$. 
It has been shown \cite{vs} that this rule adequately describes
available measured data in many cases, however, typically only spatial or temporal 
information, but not both, was taken for a particular data set.
Measurements where good statistics were collected, 
both in time and space and in a relevant region of parameter space, 
are sparse and difficult to perform. Thus, detailed 
investigation of slow motion is an urgent issue for 
future studies. 

The diffusive component of the ground motion
model considered is based on measurements of slow motion
performed at SLAC. 
First, measurements performed in the FFTB tunnel using 
the stretched wire alignment system 
over a baselength of 30~m give 
the value of $A\approx 3\cdot 10^{-7} \mu$m$^2$/(m$\cdot$s) 
on a time scale of hours \cite{fftbwire}. 
Second, a 48 hour measurement of the linac tunnel motion
performed with the linac laser alignment system
over a baselength of 1500~m gave 
$A\approx 2\cdot 10^{-6} \mu$m$^2$/(m$\cdot$s) \cite{m95}. 
Finally, recent measurements using a similar technique were made over a period of one month and show
that $A\approx 10^{-7}$--$2\cdot 10^{-6} \mu$m$^2$/(m$\cdot$s) 
for a wide frequency band of $0.01$--$10^{-6}$Hz \cite{slow1}. 
In the latter case, the major source of 
the slow $1/\omega^2$ motion  
was identified to be the temporal variations of 
atmospheric pressure coupled to spatial variations 
of ground properties \cite{slow1}.  The 
atmospheric pressure was also 
thought to be responsible for a slow variation of the parameter $A$.

\begin{figure}[th]
\vspace{-0.13cm}
\centering
{\vbox{
\epsfig{file=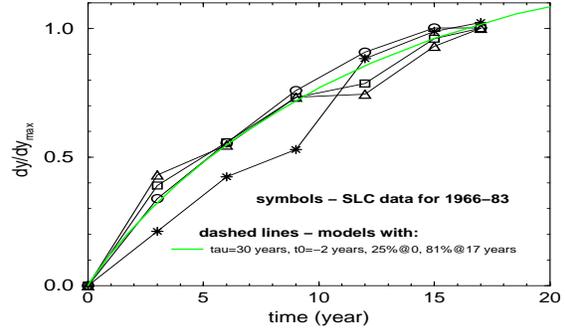,width=1.0\columnwidth,height=0.61\columnwidth}
}}
\vspace{-.7cm}
\caption{Displacement of some points of SLAC linac tunnel from 1966
through 1983 versus time and the approximation in Eq.\ (\ref{eq2}) 
with $\tau=30$ and $t_0=2$~years.}
\vspace{-.65cm}
\label{approx}
\end{figure}

The clear correlation of atmospheric pressure variation 
with deformation of the linac tunnel, observed in \cite{slow1}, 
can only be explained if one assumes some variation 
of the ground properties along the linac. This variation 
can be due to changes in the 
Young's modulus $E$, changes in the topology
such that the normal angle to the surface 
changes by $\Delta\alpha$,
or changes in the characteristic depth ${h}$
of the softer surface layers. 
A rough estimate of the tunnel deformation 
due to variation of atmospheric pressure $\Delta P$
can be expressed as 
\\[-2.5mm]
\begin{equation}
\Delta X,Y \sim \,\,\,\,  
h \frac{\Delta P}{E} \cdot \left(
\frac{\Delta E}{E}  
\,\,\,\,\,\, \mathrm{or} \,\,\,\,\,\,
\Delta\alpha
\,\,\,\,\,\, \mathrm{or} \,\,\,\,\,\,
\frac{\Delta h}{h}  \right)
\label{ptox}
\end{equation}~\\[-3.5mm]
The observed deformation of the tunnel 
$\Delta Y =50 \mu$m corresponding to $\Delta P =1000$~Pa
is consistent with this estimation if 
${\Delta}{E/E}{\sim}0.5$, ${\Delta\alpha\sim}0.5$ 
or ${\Delta}h/h{\sim}0.5$  
and if one assumes $E/h \sim 10^7$Pa/m.
The former assumption is consistent with the heterogeneous 
landscape and geology at SLAC while the latter appears to agree well
with the 
properties of the ground determined in the previous
SLAC correlation measurements,
if one assumes that $h \sim \lambda$. 
%

No direct conclusions can be drawn from 
the measurements \cite{slow1}
to determine the spatial behavior of the observed slow motion because
the relative motion was only measured for one separation distance. 
However, the topology of many 
natural surfaces (including landscapes) 
exhibits a $1/k^2$ behavior of the power spectra \cite{nature}.
Thus, it seems reasonable to expect that temporal pressure variation 
can also be a driving term of the spatial ATL-like motion. 
Furthermore, the measured parameter $A$ can be extended from 1500~m 
to a shorter scale, without contradicting 
the very short baseline measurements \cite{fftbwire} which
produced a similar value of $A$. 

It is also worth noting that the
contribution to the parameter $A$ driven by the atmosphere scales as $1/E^2$ 
or as $v_s^4$ and therefore strongly depends on geology.
Thus, the parameter $A$, at a site with a much higher
$v_s$, would not be dominated by atmospheric  
contributions, while a site with softer ground and 
a $v_s$ half that at SLAC, may have a parameter $A$ as high as 
$3\cdot 10^{-5} \mu$m$^2$/(m$\cdot$s).

\begin{figure}[b]
\vspace{-0.83cm}
\centering
{\vbox{
\epsfig{file=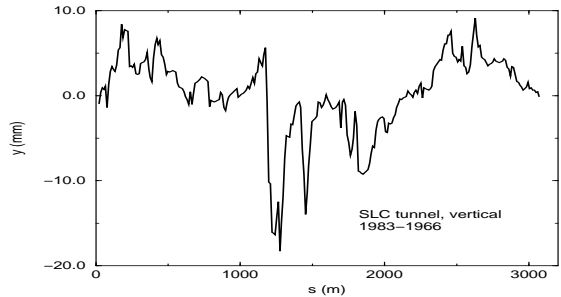,width=1.0\columnwidth,height=0.55\columnwidth}
}}
\vspace{-.68cm}
\caption{17 year motion of the SLAC linac tunnel \cite{fischer}.}
\vspace{-.05cm}
\label{sevent}
\end{figure}

Finally, very slow motion, observed on a year-to-year time scale
at SLAC, LEP, and other places, appears to be systematic in time, 
i.e.\ $\Delta X^2 \propto T^2$ \cite{pitthan}. 
For example, measurements of the SLAC linac tunnel 
between 1966 and 1983 \cite{fischer} show roughly linear motion in time 
with rates up to 1mm/year in a few locations along the
linac. Subsequent measurements indicate that the rate of this motion
has decreased over time although the direction of motion is still
similar as is illustrated in Fig.\ \ref{approx}.
In the case of SLAC, the motion may have been caused primarily by 
settling effects, while in LEP,
the cause may more likely be something different such as underground water \cite{pitthan}.

\begin{figure}[t]
\vspace{0.13cm}
\centering
{\vbox{
\epsfig{file=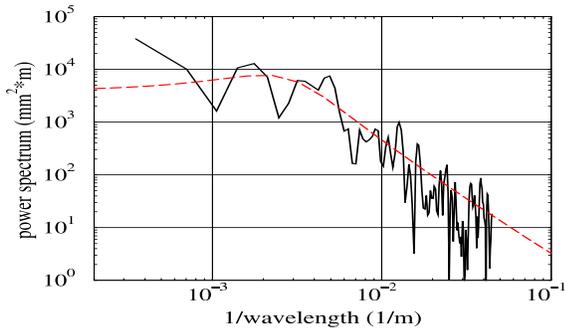,width=0.90\columnwidth,height=0.52\columnwidth}
}}
\vspace{-.35cm}
\caption{Spatial power spectrum of vertical displacements
of the SLAC tunnel for 1966 to 1983.}
\vspace{-.35cm}
\label{spatspec}
\end{figure}

The temporal dependence of earth settlement problems
typically are approximated as:
\\[-2.5mm]
\begin{equation}
\frac{\Delta y}{\Delta y_{\mathrm{max}}} \approx 1- 
\left( 1- \frac{\sqrt{{t}/{\tau}}}{(1+2\sqrt{{t}/{\tau}})}\right) 
\exp(-2.36 \, t /\tau)
\label{eq2}
\end{equation}
\\[-2.5mm]
where the typical value of $\tau$ is years. This type of 
solution exhibits $\sqrt{t}$ motion at the beginning which 
then slows and exponentially approaches $\Delta y_{\mathrm{max}}$. An example of such 
a dependence is compared with the motion observed at SLAC 
in Fig.\ \ref{approx}. One can see that the early SLAC systematic
motion can be also described reasonably well by a linear 
in time motion, though nowadays the rate of the motion 
should be already much lower. 

\begin{figure}[t]
\vspace{.225cm}
\centering
{\vbox{
\epsfig{file=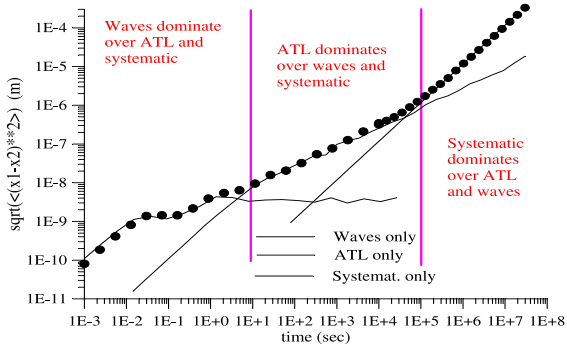,height=0.90\columnwidth,width=0.55\columnwidth,angle=-90}
}}
\vspace{-.25cm}
\caption{Rms relative motion versus time for $L=30$~m for the 
2~a.m. SLAC site ground motion model.}
\vspace{-.15cm}
\label{rmsx}
\end{figure}

The spatial characteristics of this systematic motion also seem to
follow the $1/k^2$ (or  $\Delta X^2 \propto L$) behavior. 
This is evident in the displacements of the SLAC linac \cite{fischer}
after 17 years which is shown in Fig.\ \ref{sevent}.
The corresponding spatial spectrum is shown in Fig.\ \ref{spatspec}
and it follows $1/k^2$ in the range of $\lambda$ from 20--500m.  Although there is deviation from 
the $1/k^2$ behavior at long wavelengths where there is limited data, this spectrum can be characterized as 
$P_{\mathrm{syst}}(t,k) \approx A_{\mathrm{syst}} t^2 /k^2$
with the parameter 
$A_{\mathrm{syst}} \approx 4\cdot 10^{-12} \mu$m$^2$/(m$\cdot$s$^2$)
for early SLAC.
An estimate of the rms misalignment due to this systematic motion 
is then $\Delta X^2 = A_{\mathrm{syst}} T^2 L$. 
One can see that the transition between diffusive and 
systematic motion would occur at 
$T_{\mathrm{trans}} = A/A_{\mathrm{syst}}$ which 
in our case, assuming the value 
$A=5\cdot 10^{-7} \mu$m$^2$/(m$\cdot$s) for the diffusive 
component of the SLAC ground motion model, 
would happen at about $T_{\mathrm{trans}} \approx 10^5$~s.

The SLAC ground motion model includes all of the features 
that we have described.  The transition from the `fast' to the `slow' 
motion is handled in a manner described in Ref.\ \cite{sn}. 
The absolute spectrum $p(\omega)$ and the 
spectrum of relative motion $p(\omega ,L)$ are shown in 
Fig.\ \ref{slacgm}. The systematic motion is not seen in this figure
as it corresponds to much lower frequencies.   However, it is seen in 
Fig.\ \ref{rmsx} where the rms $\Delta X$ is 
calculated for $L=30$~m by direct modeling of the ground motion using
harmonic summation \cite{sl96}.  One can see that this curve can 
be divided into three regions: wave dominated ($T \lappeq 10$~s), ATL-dominated 
($10 \lappeq T \lappeq 10^5$~s)
and systematic motion dominated ($T \gappeq T_{\mathrm{trans}} \sim 10^5$~s). 

This ground motion model is included in the PWK module of 
the final focus design and analysis code FFADA \cite{ffada}
which can perform analytical evaluations using the 
model spectra.  The model is also included 
in the linac simulation code LIAR \cite{liar} where the summation 
of harmonics is used for direct simulations of the ground motion.

\section{Conclusion}
\vspace{-.15cm}
We have presented a model of ground motion for the SLAC site. 
This model includes fast, diffusive and systematic motion with parameters that are consistent  
with the known geological structure of the 
SLAC site. It is being now used to study the performance of the various systems in the Next Linear Collider. 

We would like to thank 
C.Adolphsen,
G.Bowden,
M.Mayoud,
R.Pitthan,
R.Ruland, 
V.Shiltsev,
and 
S.Takeda
for various discussions of ground motion issues.

\end{document}